\documentclass[aps,prl,twocolumn,floatfix,superscriptaddress]{revtex4}
\usepackage{amsmath}
\usepackage{amssymb,amsfonts,amsmath}
\usepackage{pifont}
\usepackage{epstopdf}
\usepackage{multirow}
\usepackage{times}
\usepackage[T1]{fontenc}
\usepackage[latin1]{inputenc}
\usepackage{graphicx}
\usepackage{amssymb}
\usepackage{sublabel}
\usepackage{epsfig}
\newcommand{\degree}{\ensuremath{^\circ}}

\begin{document}
\title{Delayed coupling theory of vertebrate segmentation}
\date{\today}
\author{Luis G. Morelli}        \affiliation{These two authors contributed equally to this work}
                                \affiliation{Max~Planck~Institute~for~the~Physics~of~Complex~Systems,~N\"othnitzer~Stra{\ss}e~38,~01187~Dresden,~Germany}
                                \affiliation{Departamento~de~F\'{\i}sica,~FCEyN,~Universidad~de~Buenos~Aires,~Pabell\'on~I,~Ciudad~Universitaria,~Buenos~Aires,~Argentina}
\author{Sa\'ul Ares}            \affiliation{These two authors contributed equally to this work}
                                \affiliation{Max~Planck~Institute~for~the~Physics~of~Complex~Systems,~N\"othnitzer~Stra{\ss}e~38,~01187~Dresden,~Germany}
\author{Leah Herrgen}           \affiliation{Max~Planck~Institute~for~Cell~Biology~and~Genetics,~Pfotenhauerstr.~108,~01307~Dresden,~Germany}
\author{Christian Schr\"oter}   \affiliation{Max~Planck~Institute~for~Cell~Biology~and~Genetics,~Pfotenhauerstr.~108,~01307~Dresden,~Germany}
\author{Frank J\"ulicher}       \email{julicher@mpipks-dresden.mpg.de}
                                \affiliation{Max~Planck~Institute~for~the~Physics~of~Complex~Systems,~N\"othnitzer~Stra{\ss}e~38,~01187~Dresden,~Germany}
\author{Andrew C. Oates}        \email{oates@mpi-cbg.de}
                                \affiliation{Max~Planck~Institute~for~Cell~Biology~and~Genetics,~Pfotenhauerstr.~108,~01307~Dresden,~Germany}
\begin{abstract}
Rhythmic  and  sequential  subdivision  of the elongating vertebrate
embryonic  body  axis into morphological somites is controlled by an
oscillating  multicellular  genetic  network termed the segmentation
clock.  This clock operates in the presomitic mesoderm (PSM), generating
dynamic  stripe  patterns  of oscillatory gene-expression across the
field  of  PSM cells. How these spatial patterns, the clock's collective
period,  and  the underlying cellular-level interactions are related
is  not  understood.  A  theory  encompassing  temporal  and spatial
domains  of  local  and collective aspects of the system is essential to
tackle these questions. Our delayed coupling theory achieves this by
representing  the  PSM  as  an array of phase oscillators, combining
four key elements: a frequency profile of oscillators slowing across
the  PSM;  coupling  between  neighboring  oscillators; delay in
coupling;   and   a   moving   boundary  describing  embryonic  axis
elongation.  This  theory  predicts  that  the  segmentation clock's
collective  period  depends on delayed coupling. We derive an expression
for  pattern wavelength across the PSM and show how this can be used
to  fit  dynamic  wildtype  gene-expression  patterns, revealing the
quantitative  values  of parameters controlling spatial and temporal
organization  of  the  oscillators  in the system. Our theory can be
used  to  analyze  experimental  perturbations,  thereby identifying
roles of genes involved in segmentation.
\end{abstract}

The HFSP Journal, in press.

\maketitle

%
%

\section{Introduction}

\begin{figure}[h]
\begin{center}
\includegraphics[width=8.7cm]{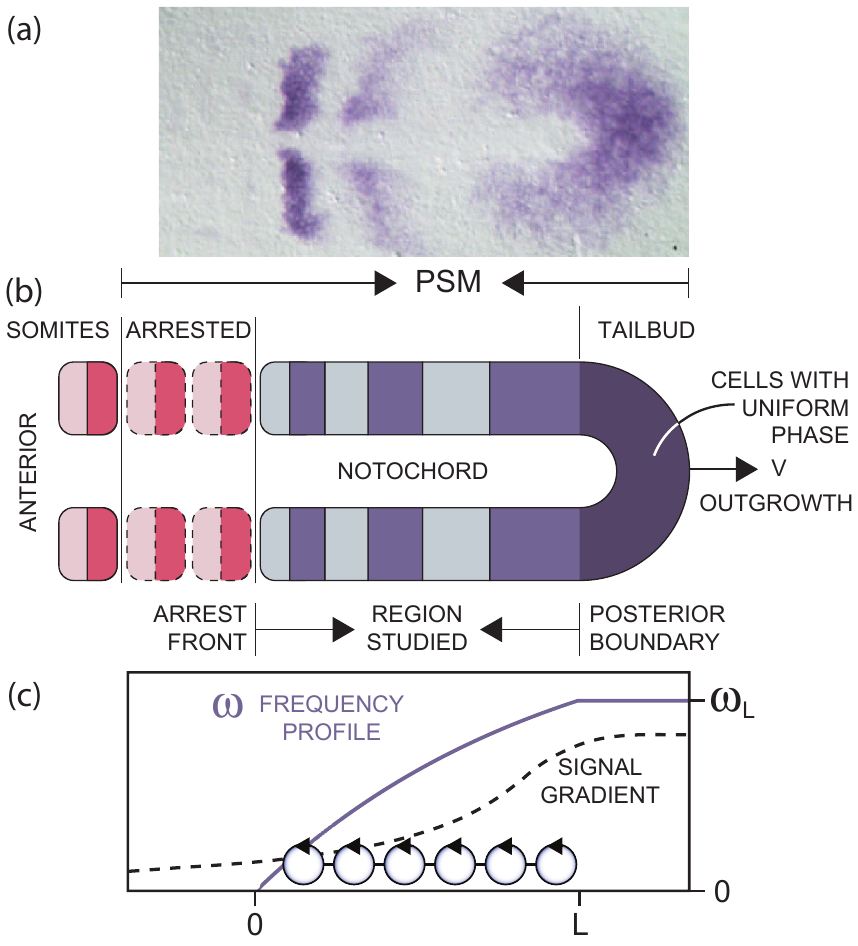}
\end{center}
\caption{\label{fig:boundaries} Representation of the pre-somitic
mesoderm (PSM). Anterior is to the left and posterior to the right.
(a) {\em In situ} hybridization \cite{oates02} showing the
expression of \emph{deltaC} mRNA in the zebrafish PSM (dorsal view).
(b) Schematic PSM together with the already determined segments
---arrested--- and the most recently formed pair of somites. The
studied region lies between the arrest front and the posterior
boundary. (c) Schematic representation of the signal gradient
spanning the PSM (broken line). The frequency profile $\omega$
related to this gradient is depicted as solid purple line, using Eq.
(\ref{eq:freq_profile}) with the width $\sigma$ given in Table I.
The length of the studied region is denoted by $L$. A linear array
of $N$ coupled oscillators is indicated.}
\end{figure}

During vertebrate development, segmentation of the continually
elongating embryonic body axis occurs rhythmically and sequentially
from head to tail in a process termed somitogenesis \cite{wolpert}.
Somites are regularly sized cell clusters that bud off periodically
from the anterior end of the posterior-most unsegmented tissue, the
pre-somitic mesoderm (PSM), with a species-specific frequency. These
transient, left-right symmetric structures are the embryonic
precursors of adult bone and muscle segments, and defects in their
formation lead to congenital birth defects \cite{bulman}. Underlying
the morphogenetic rhythm of somitogenesis, repeated waves of
oscillating gene expression sweep through the cells of the PSM from
the posterior to the anterior \cite{palm97}, see
Fig.~\ref{fig:boundaries}a and Supplementary Movie~1. These genetic
oscillations are thought to slow down and arrest at different phases
of their cycles at an anteriorly positioned arrest front that moves
in concert with embryonic elongation \cite{dubrulle01}
(Fig.~\ref{fig:boundaries}b), translating the temporal periodicity
into a striped
spatial pattern of gene expression.

Given the existence of genetic oscillators in the cells of the PSM
\cite{hirata02,masamizu06}, several questions still remain
unanswered: how does a collective segmentation period arise from the
population of individual oscillators, and what is the relation
between the overall pattern of gene expression observed in the PSM
and the oscillating expression at the cellular level. To investigate
this, we develop a theoretical description based on phase
oscillators and four key ingredients motivated by the biology: (i) a
frequency profile along the PSM, slowing down the oscillations, (ii)
coupling of oscillators, (iii) a time delay in the information
transfer between neighboring oscillators, and (iv) the existence of
a moving front that arrests the oscillations at the anterior end of
the PSM, while the posterior end moves due to embryonic outgrowth.
This delayed coupling theory provides an excellent fit to the
existing biological data, allows perturbations to the system to be
analyzed in terms of underlying processes, and predicts how
intercellular communication affects the collective period of the
segmentation clock. Below, we introduce the elements of the delayed
coupling theory.

\subsection{Phase oscillators}

To provide a simple picture of the segmentation process, Cooke and
Zeeman proposed a clock and wavefront model more than thirty years
ago \cite{cooke76}. However, to understand the role of collective
processes in the emergence of dynamic gene expression patterns in
the PSM a more detailed analysis is needed, for which methods from
other pattern forming systems can be borrowed \cite{cross93}. In
particular, the periodic expression of genes in the oscillating PSM
cells can be described at tissue level using a set of phase
oscillators, disregarding at this stage the underlying biochemical
and genetic mechanisms that generate the oscillations and their
pattern. In this phase description, each cell, or group of
synchronous cells, is represented by an oscillator, and the state of
each oscillator is characterized only by its phase in the cycle of
periodic gene expression. Oscillators with the same phase represent
cells with equivalent expression level of cyclic genes. Previous
models
described the PSM as a continuous oscillatory medium
with a phase defined at each point of the PSM, see Supplementary data of
\cite{palm97} and \cite{kaern00,jaeger01,giudicelli07,gomez08}. In this work we show
that a phase description is sufficient to compute the overall
spatiotemporal patterns of gene expression and the collective period of
the oscillations.

\subsection{Frequency profile}

It has been suggested that the arrest of the oscillations and the
observed oscillating gene expression patterns are shaped by a
spatial dependence of the frequency of the individual oscillators
\cite{palm97,kaern00,jaeger01,giudicelli07,gomez08}. A frequency
profile could be controlled by the molecular gradients observed in
the PSM, see Fig.~\ref{fig:boundaries}c, \emph{e.g.} the gradients
of the growth factor FGF
\cite{dubrulle01,sawada01,dubrulle04,wahl07} or of Wnt signaling
\cite{aulehla03,aulehla08}. Several models have recently proposed
regulatory mechanisms by which the genetic oscillations are affected
by the gradients of signaling molecules along the PSM
\cite{cinquin07,tiedemann07,santillan08,mazzitello08}.
Motivated by the changing width of the stripes of gene expression in the
PSM and the necessity that the
oscillations slow down and finally stop at the
arrest front, we include such a frequency profile in our theory. As with
the assumption of cellular oscillators, our theory does not rely on the molecular
origin of this frequency profile, and hence its inclusion is purely phenomenological.

\subsection{Coupling of oscillators}

Recent theoretical works seeking to describe spatiotemporal patterns
in somitogenesis using phase oscillators have not included coupling
between oscillators \cite{kaern00,jaeger01,giudicelli07,gomez08},
although intercellular coupling has been considered in reduced
models of regulatory circuits \cite{lewis03,cinquin03,hori06}. Here,
coupling means that oscillators can influence the phase of their
neighbors. Coupling is essential to stabilize tissue-scale patterns
against the unavoidable noise present in biological systems
\cite{jiang00,hori06,riedel07,ozbudak08}, and also to explain the
re-synchronization of surgically inverted pieces of the PSM
\cite{dubrulle01}. Thus, in this work we propose a description based
on coupled phase oscillators.

\subsection{Time delay}

Coupling between cells via signaling macromolecules, \emph{e.g.}
through the Notch pathway \cite{jiang00,hori06,riedel07,ozbudak08},
involves synthesis and trafficking of such molecules within cells.
These dynamics imply the existence of time delays, which have been
recently estimated to be in the range of tens of minutes in cell culture
\cite{heuss08}. Time delays in the coupling can have an impact on
the self-organization of coupled oscillators
\cite{schuster,niebur,yeung99,earl03,lewis03}, making their
inclusion in our theory important. For simplicity we will use here a
deterministic time delay; a more realistic description would include
a distribution of delays \cite{macdonald}.


\subsection{Moving borders due to embryonic elongation}

The embryo is a rapidly growing system, elongating about one somite
length per oscillation cycle, which takes around 25 minutes in
zebrafish, 90 minutes in chick and 120 minutes in mouse. Cells are
continuously in transit from the tailbud through the PSM, exiting it
anteriorly as somites form. Additionally, cell proliferation plays a
role during elongation, but since in the PSM it has a stochastic
character it can be considered as a potential noise source
\cite{hori06} not otherwise significantly affecting the oscillatory
dynamics \cite{zhang08}, and we do not consider it further here. To
correctly understand the formation of patterns of gene expression
and how the frequency is regulated, it is necessary to consider the
geometry and boundaries of the arena in which the process occurs.
Here we neglect changes in the antero-posterior length of the PSM or
the rate of axial growth, which occur during development at time
scales larger than the somitogenesis period
\cite{tam81,schroeter07,gomez08}. Consequently, both the arrest
front and the posterior boundary move at the same velocity $v$, see
Fig.~\ref{fig:boundaries}b.
\begin{figure}
\begin{center}
\includegraphics[width=8.7cm]{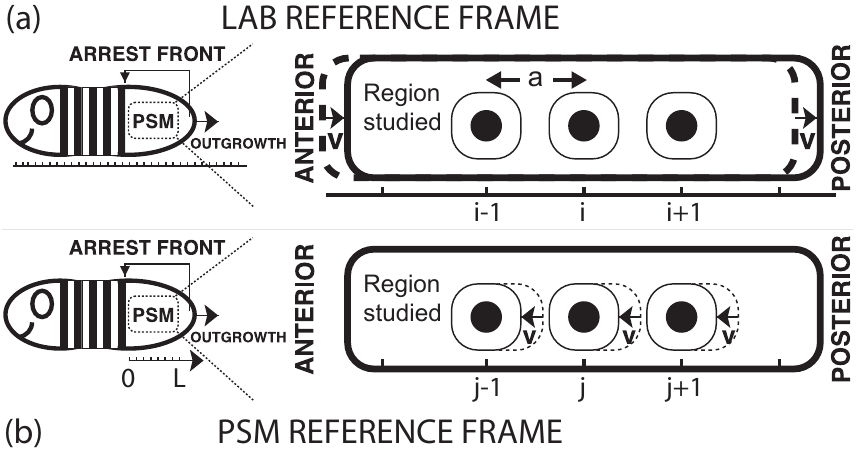}
\end{center}
\caption{\label{fig:ref-frames} Two different coordinate systems to
describe genetic oscillations in the PSM. (a) Lab reference frame:
oscillators labeled by index $i$ hold fixed positions $x_i=i a$,
where $a$ is the distance between oscillators. The PSM boundary
moves posteriorly with velocity $v$ as the embryo extends. (b) PSM
reference frame: the PSM boundaries do not move but oscillators move
through the PSM from posterior to anterior with velocity $v$.
Oscillators are constantly relabeled using symbols $j$ to denote
discrete positions relative to the arrest front. Dashed lines
indicate the state of the system at slightly earlier times.}
\end{figure}

\section{Results}

This section contains technical details of our theory. Readers
who are more interested in the basic ideas and the biological
justifications should note that a careful understanding
of the equations is not a requisite to follow the arguments we
introduce.\\

\subsection{Formulation of the delayed coupling theory}

Our model equations in the lab reference frame,
Fig.~\ref{fig:ref-frames}a, consist of a lattice of discrete phase
oscillators. This lattice comprises $N$ oscillators in the
antero-posterior direction, labeled by an index $i$. Each oscillator
occupies a position $x_i=i a$ along the PSM axis in the lab reference frame, where $a$
is the characteristic distance between oscillators, \emph{i.e.} the
average cell diameter. Hence the total physical length of the
considered system in the antero-posterior direction is $L\equiv N
a$.

As the embryo elongates with velocity $v$, the arrest front is
positioned at $vt$, where $t$ is time. Oscillators anterior to the
arrest front, $x_i \le vt$, are arrested. New oscillators are added
at the posterior boundary, situated at $vt+L$, as elongation
proceeds.
%
%
The oscillators in the studied region, with indices $vt/a\leq i < (N+vt/a)$
are weakly coupled to their $n$ nearest neighbors, denoted by the index~$k$.
In one dimension $n=2$, while in a two-dimensional square lattice $n=4$.
The phase dynamics of the coupled oscillators can be
described by
\begin{equation} \label{eq:discrete}
\dot{\theta}_i (t) = \omega_i (t) +
\frac{\varepsilon_i (t)}{na^2} \sum_{k}
\sin\left[ \theta_{k} (t-\tau_i(t))-\theta_{i} (t) \right]+\zeta_i(t),
\end{equation}
where the dot denotes time derivative, $\theta_i$ is the phase of oscillator $i$,
$\omega_i$ its intrinsic frequency, $\varepsilon_i$ the coupling
strength,
$\tau_i$ the time delay in the coupling and $\zeta_i$
is a random variable with zero average
representing different noise sources.
Our objective in this work is to characterize the basic model in the
synchronized state, and if not otherwise stated, we ignore
the effects of noise.
According to experimental evidence \cite{riedel07},
coupling strength is weak compared to other timescales in the system.
This justifies the use of the sine function in the coupling, as it is the dominant term
of any more general periodic coupling function \cite{kuramoto}.
%

To specify the shape of the frequency profile we choose, for $vt/a\leq i < (N+vt/a)$:
\begin{equation} \label{eq:freq_profile}
\omega_i(t)= \omega_\infty (1-e^{-{(i a-vt)}/{\sigma}}),
\end{equation}
where $\omega_\infty$ is a characteristic frequency scale of
individual oscillators, and  $\sigma$ is a measure of the
characteristic distance over which the frequency profile decreases
from high to low values (see Fig.~\ref{fig:boundaries}c). Below we
will show that our choice of Eq.~(\ref{eq:freq_profile}) is
consistent with experimental observations, and we will determine
$\sigma$ from the width of the stripes of gene expression in the
PSM. Qualitatively similar choices for the frequency profile have
been used before \cite{kaern00,jaeger01}. For simplicity we have
chosen the frequency to be strictly zero at the arrest front. Note
however that this is not biologically necessary: a very low value of
the frequency at the arrest front means a very large period of
oscillation. If this period is much larger than any other time scale
involved in the process, it determines in practice an arrested oscillation,
which can specify the downstream fixed pattern that
eventually sets the position of somite boundaries. Furthermore,
since the oscillations can be coupled to a bistable system arising
from opposing signaling gradients in the PSM, long period
oscillations at the arrest front could be stopped by a bistable
transition \cite{goldbeter07,santillan08}.

For convenience we introduce the parameter
$\omega_L\equiv\omega_\infty (1-e^{-{L}/{\sigma}})$, that represents
the intrinsic frequency of the oscillators at the posterior boundary
of the system. Based on in situ experiments that show a largely
uniform spatial expression of cyclic genes in the tailbud at any
given stage of the cycle (Fig.~\ref{fig:boundaries}a), we define a
uniform phase (Fig.~\ref{fig:boundaries}b) and homogeneous frequency
$\omega_L$ (Fig.~\ref{fig:boundaries}c) in the tailbud, posterior to
the notochord and the region we study here. This homogeneity would
be favored by the strong cell mixing \cite{mara07}, and high,
potentially saturating uniform levels of the signaling molecules
that establish the gradient anterior to this region
\cite{sawada01,dubrulle01,dubrulle04,wahl07,aulehla03,aulehla08}.
Furthermore, the shape of $\omega_i$ described by
Eq.~(\ref{eq:freq_profile}) resembles the posterior branch of FGF
receptor saturation proposed in \cite{goldbeter07}.

The coupling could also be position dependent. In particular, since
the oscillators anterior to the arrest front stop cycling, they can
not influence the active oscillators in the interval $vt/a\leq i <
(N+vt/a)$. We take this into account imposing
$\varepsilon_{i}=\varepsilon_{0}=0$ for $i <vt/a$. For simplicity,
in this work we consider the coupling strength
$\varepsilon_{i}\equiv\varepsilon$ and the time delay
$\tau_i\equiv\tau$ to be constant posterior to the arrest front.

\subsection{Numerical simulations in two dimensions}

\begin{figure}
\begin{center}
\includegraphics[width=8.7cm]{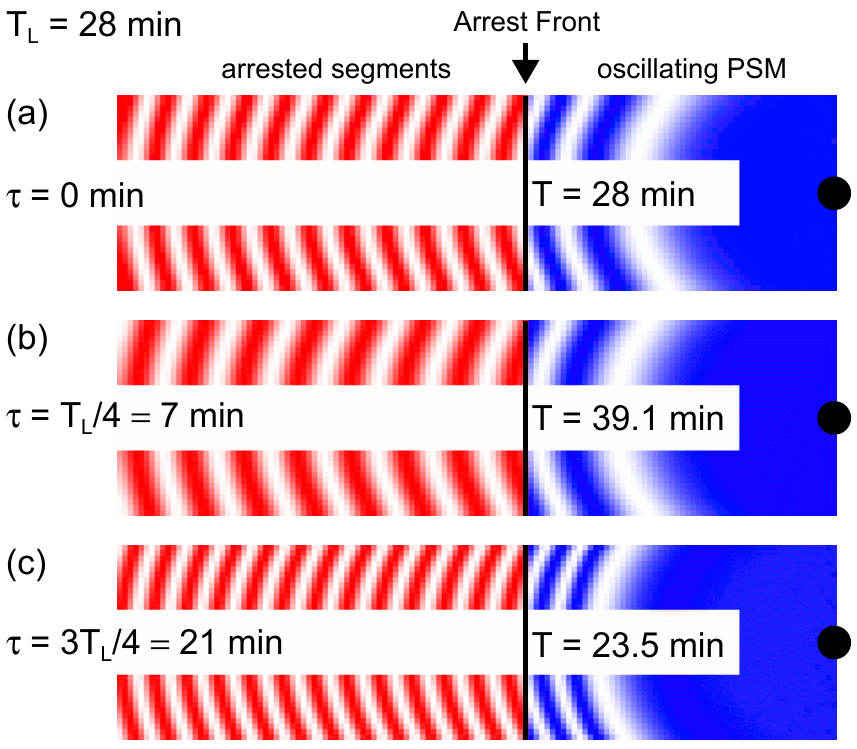}
\end{center}
\caption{\label{fig:2D_simulations} Numerical simulation of
segmentation using Eq.~(\ref{eq:discrete}) in a growing
two-dimensional geometry. Color intensity indicates the value of
$\sin\theta$ of the phase $\theta$: white is $\sin\theta=1$ and dark
(red or blue) is $\sin\theta=-1$. Vertical line indicates the
position of the arrest front, with the oscillating PSM to its right
(blue) and the arrested pattern to the left (red). Intrinsic
frequency is a decaying function of the distance to the black dot at
the posterior boundary, causing the curvature of the stripes. We
have used parameters determined for zebrafish at 28\degree C, see
Table I, and $\omega_L = 0.224$min$^{-1}$, see main text for
details. We have chosen to display three illustrative values of the
time delay (a) $\tau=0$ min, (b) $\tau=T_L/4\approx7$ min and (c)
$\tau=3T_L/4\approx21$ min. Delayed coupling affects the global
frequency of oscillations according to
Eq.~(\ref{eq:Omega-epsilon-tau}). The stripes of the oscillating PSM
pattern and the segment length of the frozen pattern change
accordingly. Movies available as supplementary material.}
\end{figure}

To gain insight into the role of delayed coupling in setting the
period and the pattern of the genetic oscillations, as well as to
illustrate the formation of realistic wave patterns within our
theory, we performed computer simulations of Eq.~(\ref{eq:discrete}) in a
two dimensional geometry using different values of the time delay
(Fig.~\ref{fig:2D_simulations} and Supplementary Movies~2, 3 and 4).
Although the theory represents generic vertebrate segmentation, here
we use parameters from the zebrafish embryo, Table I. For the
intrinsic frequency at the posterior we chose
$\omega_L=0.224$~min$^{-1}$, corresponding to an intrinsic period of
$T_L=2\pi/\omega_L=28$~min. We show simulations with no time delay
($\tau=0$~min), a short delay compared to the intrinsic period
($\tau=T_L/4=7$~min), and a long delay close to the intrinsic period
($\tau=3T_L/4=21$~min). The latter is consistent with the
experimental observation of  tens of minutes for intercellular
communication times \cite{heuss08}.

Fig.~\ref{fig:2D_simulations}a shows a snapshot of the simulation
with no time delay. Not surprisingly, the collective period ---the time
needed to form one new arrested segment, and also the time after
which the oscillating pattern in the PSM repeats itself--- is
unchanged with respect to the intrinsic period at the posterior. The
arrested segments have a length of $S\approx 7$~cell diameters. The
case of short delay with respect to the period, $\tau=T_L/4$,
qualitatively represents the situation in species with a relatively
long segmentation period, such as mouse.
Fig.~\ref{fig:2D_simulations}b shows that in this case, the effect
of the delay in coupling is to slow down the collective period ($T=39.1$~min)
with respect to the intrinsic period ($T_L=28$~min). Further, the
arrested segments are longer ($S\approx 10$~cell diameters) than in
the case without delay, and there is a smaller number of gene
expression stripes in the PSM, with increased size.
Surprisingly, when the delay
is made longer, the trends observed with the short delay are
inverted. Fig.~\ref{fig:2D_simulations}c shows that for the time
delay of $\tau=3T_L/4$, the collective period ($T=23.5$~min) is shorter than
the intrinsic period. Moreover, the arrested segments are also
shorter ($S\approx 6$ cell diameters) than in both previous cases,
as are the stripes of  oscillating gene expression in the PSM.

This puzzling results show that delayed coupling introduces
non-trivial effects to the system, large enough to be observable in
real experiments. In order to understand these effects, in the following
we perform an analysis of Eq.~(\ref{eq:discrete}), studying first
the emergence of the collective period from the parameters of the theory
and then turning our attention to the spatial pattern. For this
purpose, we will write Eq.~(\ref{eq:discrete}) in a more convenient
manner.

\subsection{PSM reference frame}

It is useful to consider the dynamics in the PSM reference frame,
Fig.~\ref{fig:ref-frames}b, where the oscillations can be
characterized by a stationary phase profile and a collective frequency.
For simplicity from here on we use a one dimensional description of
the system, with $n=2$. In the lab reference frame
(Fig.~\ref{fig:ref-frames}a) the symbol $i$ represents a fixed
oscillator. In the PSM reference frame (Fig.~\ref{fig:ref-frames}b)
we introduce the symbol $j$ to label fixed discrete positions
relative to the arrest front. The label $j$ runs from $j=0$ at the
arrest front to $j=N$ at the posterior boundary of the PSM. Discrete
position $j$ is occupied by different oscillators as the system
evolves in time. For convenience we have included in the description
the last arrested oscillator, $j=0$.

In the PSM reference frame, the frequency profile is stationary,
$\omega_j=\omega_\infty (1-e^{-{j a}/{\sigma}})$. Re-expressing
Eq.~(\ref{eq:discrete}) in this PSM reference frame, an extra term
describes the drift of the phase due to the movement of the cells
relative to the PSM boundaries. The resulting phase dynamics are
given by:
\begin{eqnarray} \label{eq:discrete-psm}
\dot{\varphi}_j(t) = \omega_j +v[\varphi_{j+1}(t)-\varphi_j(t)]+\\
\nonumber \frac{\varepsilon}{2a^2} \!\! \sum_{k=j+p\pm 1} \!\!\!\!
\sin\left[ \varphi_{k} (t-\tau)-\varphi_{j} (t) \right].
\end{eqnarray}
Here $\varphi_j$ is the phase at position $j$ relative to the arrest front and
$p=[v\tau/a]$ is the nearest integer to $v\tau/a$,
representing the distance a cell moves during the
time it takes for a signal from a neighbor to arrive.
Note that now the coupling
is non-local: due to the time delay and cell movement, the
neighbors of an oscillator with position $j$ had positions $j+p+1$ and
$j+p-1$ at the time the signal was sent.
\begin{figure}[t]
\begin{center}
\includegraphics[width=8.7cm]{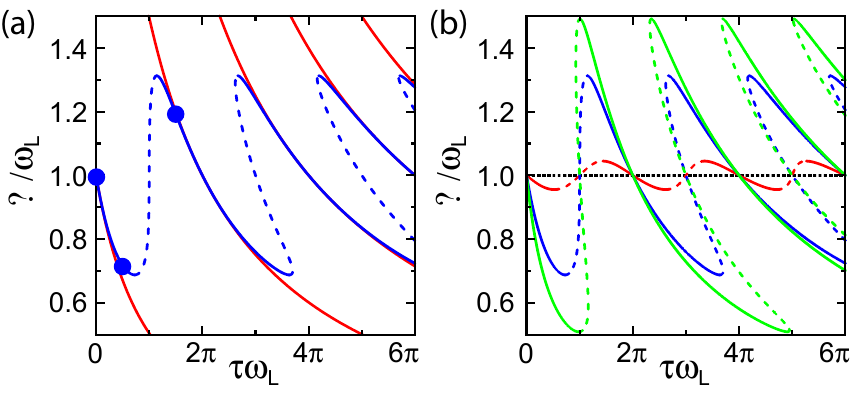}
\end{center}
\caption{\label{fig:omega.tau} Global frequency $\Omega$ of
somitogenesis as a function of time delay and coupling strength. (a)
Dimensionless global frequency $\Omega/\omega_L$ and time delay
$\tau\omega_L$ are displayed for constant coupling $\varepsilon =
0.07$ (cell diameter)$^2$min$^{-1}$, and intrinsic frequency
$\omega_L=0.224$ min$^{-1}$. Analytical solutions of
Eq.~(\ref{eq:Omega-epsilon-tau}) shown as blue lines: solid lines
stable solutions of Eq.~(\ref{eq:Omega-epsilon-tau}), dashed lines
unstable solutions of Eq.~(\ref{eq:Omega-epsilon-tau}). Blue dots
correspond to numerical integration of the discrete model in two
dimensions as given by Eq.~(\ref{eq:discrete}), for the two cases
illustrated in Fig.~\ref{fig:2D_simulations}c,d. Red lines: global
frequency as a function of delay in the continuum limit, showing its
ranges of validity. (b) Global frequency $\Omega$ as a function of
time delay for different coupling strengths obtained from the
solution of Eq.~(\ref{eq:Omega-epsilon-tau}) with $\varepsilon =
0.11$ (cell diameter)$^2$min$^{-1}$ (green), $\varepsilon = 0.07$
(cell diameter)$^2$min$^{-1}$ (blue) and $\varepsilon = 0.03$ (cell
diameter)$^2$min$^{-1}$ (red). Solid lines are stable solutions,
dashed lines are unstable solutions. Dotted line at
$\Omega/\omega_L=1$ corresponds to vanishing coupling, $\varepsilon
= 0$ (cell diameter)$^2$min$^{-1}$.}
\end{figure}

\subsection{Steady state ansatz and collective frequency}

The oscillating gene expression pattern in the PSM repeats after a full
period $T=2\pi/\Omega$ of oscillation \cite{palm97,masamizu06}, where $\Omega$
is the collective frequency of the oscillation.
This leads to
the steady state ansatz $\varphi_j (t) = \Omega t + \phi_j$, where
$\phi_j$ is the stationary phase profile
describing the pattern
in the PSM.
With this ansatz we obtain
from Eq.~(\ref{eq:discrete-psm}):
\begin{equation} \label{eq:discrete-psm-spatial}
\Omega = \omega_j +v[\phi_{j+1}-\phi_j]
+ \frac{\varepsilon}{2a^2} \!\!
\sum_{k=j+p\pm 1} \!\!\!\!\! \sin\left[ \phi_{k}-\phi_{j}
-\Omega\tau \right].
\end{equation}
The collective frequency of oscillations $\Omega$ is equivalent to the
rate of somite formation. Note that the instantaneous frequencies $\dot{\theta}_i$
of individual oscillators depend on position and are in general
different from $\Omega$.

\subsection{Anterior boundary condition sets the segment length}

To determine the collective frequency $\Omega$ we need to specify the
boundary conditions, namely the conditions that $\phi_j$ fulfills at
the borders of the studied region ($j=0$ and $j=N$,
Fig.~\ref{fig:boundaries}b). This boundary should not be confused
with somite boundaries, which we do not discuss in this paper.

At the arrest front ($j=0$), the fact that $\omega_0=0$ and
$\varepsilon_0=0$ implies with Eq.~(\ref{eq:discrete-psm-spatial})
that $(\phi_1-\phi_0)=\Omega/v=2\pi/vT$. Thus, the anterior boundary
condition determines the wavelength of the arrested pattern, which
is the segment length $S=2\pi/(\phi_1-\phi_0)=vT$: the segment
length is the distance advanced by the arrest front during one
oscillation period \cite{cooke76}.

\subsection{Coupling and delay affect the collective period}

At the posterior boundary of the PSM, we assume that new cells are
added into the system with phase $\phi_N$. To implement this we
impose in Eq.~(\ref{eq:discrete-psm-spatial}) $p$ boundary
conditions, $\phi_j= \phi_N$ for $j=N+1,\ldots,(N+p)$, accounting in
this way for the effective non-locality of the coupling. We base
this choice on the experimental observation of cyclic gene mRNA
patterns, which maintain a smooth expression profile, and hence
approximately homogeneous phase, across the interface between
tailbud and posterior PSM (e.g. Fig.~\ref{fig:boundaries}a).

Substituting the posterior boundary condition $\phi_{N+1}=\phi_N$ in
Eq.~(\ref{eq:discrete-psm-spatial}) we
obtain a relation for the collective frequency of oscillations
(see also \cite{schuster,niebur,yeung99,earl03}):
\begin{equation} \label{eq:Omega-epsilon-tau}
\Omega = \omega_L - \varepsilon\sin\left(\Omega\tau \right).
\end{equation}
The solutions to this equation are shown in Figs.
\ref{fig:omega.tau}a and \ref{fig:omega.tau}b. Results from
numerical simulations of Eq.~(\ref{eq:discrete}) in two spatial
dimensions show that the collective frequency indeed fulfills
Eq.~(\ref{eq:Omega-epsilon-tau}), see blue dots in
Fig.~\ref{fig:omega.tau}a.


For a given set of parameters $\omega_L$, $\varepsilon$, and $\tau$,
Eq.~(\ref{eq:Omega-epsilon-tau}) allows for multiple solutions for
the collective frequency $\Omega$. Independent measurement of coupling
strength $\varepsilon$, collective frequency $\omega_L$ and collective
frequency $\Omega$ would allow the determination of possible values
of the delay $\tau$ consistent with
Eq.~(\ref{eq:Omega-epsilon-tau}). Experimentally,
this can be done studying situations where the intrinsic
cellular oscillations are altered (modified $\omega_L$) or where the
coupling strength is altered (modified $\varepsilon$) and using the
observed values of $\Omega$ to fit $\tau$.

A linear stability analysis following \cite{yeung99,earl03} reveals
that when $\cos(\Omega\tau)>0$ the solution to
Eq.~(\ref{eq:Omega-epsilon-tau}) is stable, and unstable otherwise,
see continuous and dashed lines in Fig.~\ref{fig:omega.tau}.
Consequently, multi-stability occurs for large values of
$\varepsilon$ and $\tau$. As seen in Fig.~\ref{fig:omega.tau}a, for
the biologically plausible parameters that we use in the figure,
there is a small gap of time delay values between the first and the
second stable branches of the solution. This happens around
$\tau\omega_L=\pi$, which means that the delay is close to half the
intrinsic period of the cellular oscillators in the posterior PSM,
$\tau\approx T_L/2\equiv \pi/\omega_L$. For the parameters in
Fig.~\ref{fig:omega.tau}a, larger values of the delay always involve
at least one stable solution. Note that values of the collective
frequency equal to the intrinsic frequency, $\Omega=\omega_L$, are
only possible for delays equal to integer and semi-integer multiples
of the intrinsic period $T_L$: these solutions are stable in the
case of integer multiples ($\tau=$integer$\times T_L$) and unstable
in the case of semi-integer multiples ($\tau=($integer$+1/2)\times
T_L$). However, stable solutions are possible for these latter
delays, albeit with the collective frequency $\Omega$ different than the
intrinsic frequency $\omega_L$.

Eq.~(\ref{eq:Omega-epsilon-tau}) provides an explanation for the
non-monotonic behavior of the collective period observed in
Fig.~\ref{fig:2D_simulations}. Moreover, the simulation results
coincide quantitatively with the prediction of
Eq.~(\ref{eq:Omega-epsilon-tau}), as shown by the three dots in
Fig.~\ref{fig:omega.tau}a. Eq.~(\ref{eq:Omega-epsilon-tau}) is
biologically relevant: the collective frequency or period of
somitogenesis emerges as a self-organized property, and depends not only
on the intrinsic frequency of individual cells, but also on the
coupling strength and the time delay (Fig.~\ref{fig:omega.tau}).
Note that $\Omega$ does not depend on the specific shape of the
frequency profile, and the period is set by the uniform phase cell
population in the tail, which is the pacemaker of the whole
oscillatory process.
\begin{figure}[t]
\begin{center}
\includegraphics[width=8.7cm]{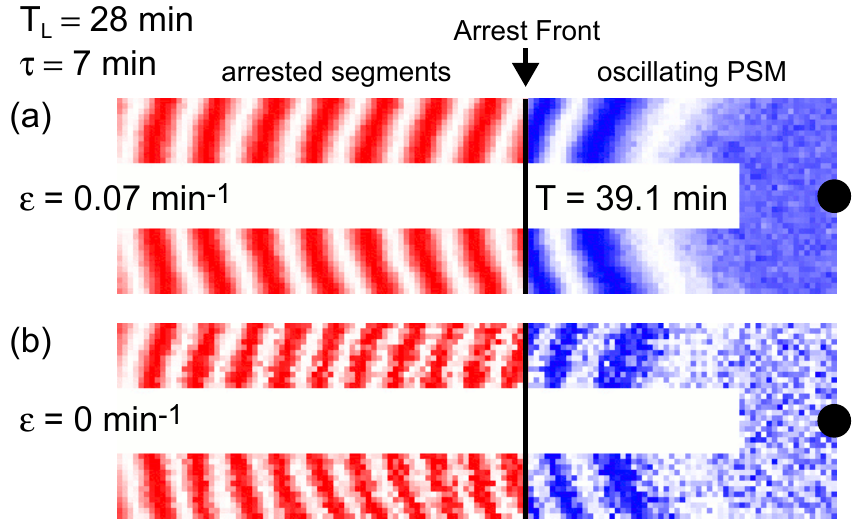}
\end{center}
\caption{\label{fig:noise.notch} Effects of noise in the delayed
coupling theory. We include a white Gaussian noise as discussed in
the text. (a) Delayed coupling is robust against the influence of
noise. Parameters as in Fig.~\ref{fig:2D_simulations}b. (b) Impaired
coupling results in segmentations defects. After initial
synchronization with resulting segments not shown, coupling is
turned off ($\varepsilon=0$ (cell diameter)$^2$min$^{-1}$). The
first segments have recognizable boundaries, but posterior segments
are increasingly disrupted due to the effect of noise. Parameters as
in (a).}
\end{figure}
\begin{figure}[t]
\begin{center}
\includegraphics[width=8.7cm]{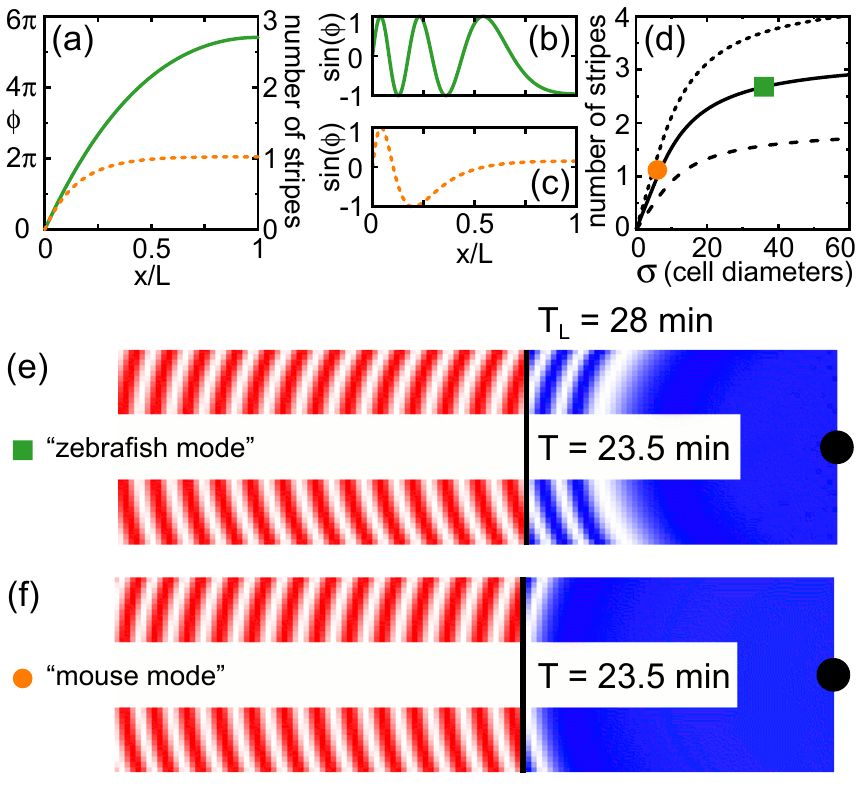}
\end{center}
\caption{\label{fig:phase.profile} Phase profile in the PSM in the
continuum limit. (a) Phase profile as a function of relative
position, given by Eq.~(\ref{eq:solution.phase.nondimensional}).
Left axis: phase relative to the arrest front. Right axis:
corresponding number of gene expression stripes. The green solid
line corresponds to the set of parameters obtained from zebrafish
data, see Table I, using $T_L=28$ min and $\tau=21$ min for
illustration. Orange dotted line corresponds to $\sigma=6$ cell
diameters. (b,c) Waveform of the expression pattern represented as
$\sin \phi$. (d) Number of stripes in the PSM as a function of
$\sigma$ from Eq.~(\ref{eq:number-stripes}), where $\sigma$ is the
parameter describing the decay length of the frequency profile.
Black solid line corresponds to parameters in Table I, with $\sigma$
variable. Green square dot: $\sigma$ obtained from zebrafish data,
see Table I. Orange circular dot: mouse mode, see orange dotted
curve in (a,c). Dotted and dashed curves correspond to higher and
lower values of global frequency, which can potentially be affected
by the intrinsic frequency, the coupling strength, or the time
delay, see Eq.~(\ref{eq:Omega-epsilon-tau}). (e) Zebrafish mode,
reproducing panel (c) of Fig.~\ref{fig:2D_simulations} for
comparison with panel (f). (f) Mouse/chick mode: zebrafish
parameters as in (e), but with a sharper frequency profile, $\sigma
= 6$ cell diameters. Only one wave of expression appears in the PSM,
in contrast to the almost three waves in (e). Movies available as
supplementary material.}
\end{figure}

\subsection{Delayed coupling keeps the oscillations synchronized}

We have seen how the presence of time delay in the coupling can have
an important effect in the spatiotemporal patterns of gene
expression. A critical biological function of intercellular
coupling is to keep neighboring cells oscillating in
synchrony \cite{jiang00,hori06,riedel07,ozbudak08}. To
demonstrate that delays in the coupling allow this function, and to
and to showcase the role of noise in our
theory, we simulate the phenotype of a class of mutant embryos in
which coupling is strongly reduced \cite{jiang00,riedel07}. To do
this we include an additive white Gaussian noise in the simulations,
with zero mean and correlations $\langle \zeta_i(t)\zeta_k(t')
\rangle = 2Q\delta(t-t')\delta_{ik}$, and choose $\sqrt{2Q}=0.036$~min$^{-1}$
for illustrative purposes. With coupling as in Table I,
the process is not disrupted by noise, Fig.~\ref{fig:noise.notch}a
and Supplementary Movie~5. When coupling is disrupted, the
simulation exhibits posterior segmentation defects,
Fig.~\ref{fig:noise.notch}b and Supplementary Movie~6, resembling
Delta-Notch mutant phenotypes in zebrafish
\cite{holley00,holley02,itoh03,julich05,oates05}. A more subtle
feature of Fig.~\ref{fig:noise.notch}b is the change in segment
length after coupling has been disengaged. As soon as the coupling
is removed, the effects of time delays are no longer present, and cells
can oscillate at their intrinsic frequencies. Because at the time
the coupling is turned off there is an established pattern in the
oscillating PSM, it takes a few cycles for this information to be
wiped out and to reach the new steady state value of the segment
length.

Although out of the scope of this work, our model provides a simple
framework to study the effects of different kinds of noise on
segmentation. This interesting possibility remains open for future
work.

\subsection{Spatial patterns of gene expression}

While the collective frequency $\Omega$ describes the temporal
regularity of somitogenesis, the spatial pattern of gene expression
in the PSM is characterized by the phase profile $\phi_j$. To
evaluate the phase profile it is convenient to introduce a continuum
limit where the spatial coordinate takes continuous values, denoted
by $x$, replacing the discrete index $j$, see Methods. The
stationary phase profile $\phi(x)$, see
Fig.~\ref{fig:phase.profile}a, can be compared to quantitative
experimental measurements of the pattern, such as the width of the
stripes of gene expression reported in \cite{giudicelli07}. We
define the wavelength $\lambda$ as the distance of two points in the
PSM with a phase difference of $2\pi$, see
Fig.~\ref{fig:lewis.fit}a. The wavelength is large close to the tail
and becomes smaller close to the arrest front where it matches the
segment length. Using the continuum formalism we find an expression
for the dependence of $\lambda$ with the position $x$ of the
stripe's center relative to the arrest front
\begin{equation}\label{eq:wavelength2}
x\approx\sigma\log\left[\frac{\sinh\left({\lambda}/{2\sigma}\right)}{{\pi}{\nu^{-1}}{(1+\eta)^{-1}}+({\lambda}/{2\sigma})\,e^{-L/\sigma}}
\right].
\end{equation}
Here $\nu$ and $\eta$ are dimensionless parameters relating
intrinsic frequency, coupling, time delay, elongation speed and the
frequency profile, as defined in the Methods. In
Fig.~\ref{fig:lewis.fit}b we show the fit of
Eq.~(\ref{eq:wavelength2}) to the wavelengths obtained from the raw
data in \cite{giudicelli07}: distances between consecutive points
with equal level of {\em her1} expression in zebrafish embryos
around the ten somite stage and raised at 28\degree C. The equation
fits very well to the data, showing that our choice of
Eq.~(\ref{eq:freq_profile}) for the frequency profile is consistent
with observations.
\begin{table}[t]
\begin{center}
\begin{tabular}{|c|l|l|}
\hline
$T$ & Period of somite formation \cite{schroeter07} & 23.5 min \\
\hline
$\Omega$ & Global frequency, $2\pi/T$ & 0.267 min$^{-1}$ \\
\hline
$S$ & Somite size (our own experimental estimation) & 6 cd  \\
\hline
$v$ & Velocity of the arrest front, $v=S/T$ & 0.255 cd/min\\
\hline
$ \varepsilon $ & Coupling strength \cite{riedel07} & 0.07 cd$^2$min$^{-1}$  \\
\hline \hline
$L $ & PSM length  & 39 cd  \\
\hline
$\sigma$ & Decay length of the frequency profile & 36 cd \\
\hline
$\tau$ & Time delay   & \multirow{2}{*}{Not determined}\\
\cline{1-2}
$\omega_L$ & Intrinsic freq. in the posterior PSM & \\
\hline
\end{tabular}
\end{center}
\caption{Parameters of the delayed coupling theory and their values
in zebrafish embryo at 28\degree C near the ten somite stage. The
first five parameters have been determined before or come from our
observations. PSM length $L$ and decay length of the frequency
profile $\sigma$ come from fits of our theory to experimental data
in \cite{giudicelli07}. The parameters $\omega_L$ and $\tau$ could
not be determined independently in this work, but they are related
by Eq.~(\ref{eq:Omega-epsilon-tau}). Note that parameters change
with temperature and throughout development \cite{schroeter07}.
We choose as length unit one cell diameter (cd), in terms of which
the distance between neighbor oscillators is $a=1$cd.}
\label{table:param}
\end{table}

\subsection{Parameter values}
\label{sec:param}

\begin{figure}[t]
\begin{center}
\includegraphics[width=8.7cm]{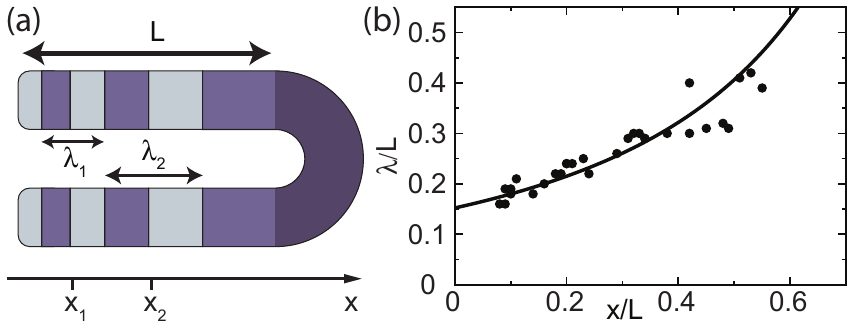}
\end{center}
\caption{\label{fig:lewis.fit} Wavelength of the pattern as a
function of the position $x$ in the PSM, where $L$ is the length of
the part of the PSM considered. (a) Schematic representation of the
wavelength $\lambda$. (b) Fit of Eq.~(\ref{eq:wavelength2}) to the
experimental data obtained from wildtype zebrafish in
\cite{giudicelli07}. Best fit parameters are $\mu = 1.08$ and
$\nu(1+\eta)=57.8$. ($\mu$, $\nu$ and $\eta$ are dimensionless
quantities defined in the Methods.)}
\end{figure}

From the fit to data obtained from wildtype zebrafish shown in
Fig.~\ref{fig:lewis.fit}b we determine $L/\sigma=1.08$ and
$\nu(1+\eta)=57.8$. We estimate the parameters $L$ and $\sigma$
using the definitions of $\nu$ and $\eta$ and the measured values of
$T$ and $S$, see Table I. Time delay affects both the collective
frequency and the wavelength of the gene expression patterns. As we
show in the Methods section, delayed coupling introduces a
renormalization of both frequency and coupling strength. The effects
of the time delay are thus included in the dimensionless
renormalized parameters of Eq. (6), but the fit of spatial gene
expression patterns does not allow the separation of the
contribution of the time delay from that of the intrinsic frequency,
and hence these two parameters remain undetermined from this fit.
The intrinsic frequency at the posterior $\omega_L$, and the time
delay $\tau$, are related through Eq.~(\ref{eq:Omega-epsilon-tau}).
Thus experimental determination of one would suffice to calculate
the other if the coupling strength and collective frequency are known.

From our estimated parameters in Table~I the value of the frequency
$\omega_L$ can be up to 30\% higher or lower than the collective
frequency $\Omega$, see Fig. \ref{fig:omega.tau}a. For an intrinsic
period $T_L=2\pi/\omega_L$ around 28~min, this implies that changing
coupling strength or delay time could situate the collective period in a
range between 21 and 40~min, in qualitative agreement with the
magnitude of period change from simulations of the genetic
regulatory network model in \cite{lewis03} for two coupled cells
\cite{leier08}. Note that this period change is only possible due to
the presence of delays in the coupling, both in our theory (see
Eq.~(\ref{eq:Omega-epsilon-tau})) and in the model in
\cite{leier08}. This difference in period is large and should be
accessible to experimental observation, allowing at the same time
for numerical determination of the values of the time delay $\tau$
and the intrinsic frequency $\omega_L$.

\section{Discussion}

We have constructed a phenomenological theory describing the tissue-level dynamics of
the vertebrate segmentation clock employing phase oscillators to
represent cyclic gene expression in the cells of the PSM. As key
ingredients of the theory, we considered (i) the existence of a
frequency profile, (ii) coupling between oscillators, (iii) time
delay in this coupling, and (iv) moving boundaries corresponding to
embryonic elongation and the moving arrest front. Although these
four elements have been considered before, here we combine them in a
unified framework. In this theory, tissue-level phenomena are
generated by the interaction of cellular properties. For example,
the collective frequency of oscillation of the PSM, related to the
segmentation rate, depends on the intrinsic frequency at the
posterior, the coupling strength and the time delay in the coupling,
Eq.~(\ref{eq:Omega-epsilon-tau}); the spatial wavelength of gene
expression stripes in addition depends on the shape of the frequency
profile. Knowledge of the molecular underpinnings is not necessary
for this mesoscopic description. By fitting the phase profiles
obtained in our continuum limit to the existing data from a
vertebrate embryo, we obtained a description of the tissue- and
cellular-level processes controlling period and pattern in the
system that is both quantitative and predictive. This framework can
now be used to analyze experimental and evolutionary variants of
embryonic segmentation or other permutations of growing, oscillating
systems.

Note that the basic relationship of a clock and wavefront type model
for embryonic segmentation, as initially proposed by Cooke and
Zeeman \cite{cooke76}, is that the length of a segment is the
product of the arrest wavefront velocity and the period of the
clock. In our description the population of oscillators create a
pattern with a collective frequency, that together with the movement of
the arrest front gives rise to a segment length consistent with the
clock and wavefront picture.

\subsection{Variation of stripe patterns for different animal species}

We have compared our theory to zebrafish data, but it applies
equally well to other vertebrate species, since it does not involve
species-specific details. The difference between what is termed a
zebrafish mode of oscillation in somitogenesis and a mouse/chick
mode, observed also in medaka \cite{gajewski06}, can be
characterized as follows: in the zebrafish mode, several waves of
gene expression sweep simultaneously through the PSM, \emph{i.e.}
multiple stripes of expression are detected in {\em in situ}
experiments; in mouse/chick mode, only one wave is observed. The
zebrafish mode applies also to snakes, where up to nine waves of
gene expression have been observed \cite{gomez08}. Within our theory
these different modes are characterized by the phase difference between the
arrest front and the posterior border: the number of stripes of gene
expression in the PSM is $(\phi(L)-\phi(0))/2\pi$, see
Figs.~\ref{fig:phase.profile}a,b,c. From
Eq.~(\ref{eq:solution.phase.nondimensional}) in the Methods we find:
\begin{equation}\label{eq:number-stripes}
\text{Number of stripes}\approx\frac{\sigma}{vT}-\frac{L}{(e^{L/\sigma}-1)vT}=\frac{1}{\mu s}-\frac{1}{(e^{\mu}-1)s}.
\end{equation}
This expression can be written as a function of only two
dimensionless parameters: the ratio $\mu=L/\sigma$ between the
system length and the decay length of the frequency profile and the
ratio $s=S/L=vT/L$ between the segment length and the system length.
The number of stripes is a decreasing function of both these ratios:
smooth frequency profiles with long decay lengths, as well as small
segment lengths, favor a large number of stripes of gene expression,
as in the zebrafish mode, see Fig.~\ref{fig:phase.profile}d. The
coupling strength and time delay do not appear in
Eq.~(\ref{eq:number-stripes}) because we have neglected for
simplicity higher order terms in $\varepsilon$ where they show up
explicitly. Note however that the collective period $T=2\pi/\Omega$ in
Eq.~(\ref{eq:number-stripes}) does depend on both the coupling
strength and the time delay through
Eq.~(\ref{eq:Omega-epsilon-tau}), see Fig.~\ref{fig:phase.profile}d.
Thus, in the same way that it may modify the collective period (as
discussed in section~\ref{sec:param}), the effect of delayed
coupling can vary up to 30\% the number of stripes of gene
expression observed in the PSM compared to a system without
coupling, see Fig.~\ref{fig:2D_simulations}.

A similar formalism for calculating the number of stripes has
recently been published as Supplementary Material in \cite{gomez08}.
The underlying theory was previously proposed in
\cite{kaern00,jaeger01}, and is the same as our continuum theory,
but without coupling, and hence without the effects caused by the
delay in the coupling. However, in \cite{gomez08} no explicit choice
for the shape of the frequency profile is made, hence the resulting
formula for the number of stripes is a function of an unknown
integral, rather than a closed formula. Our
Eq.~(\ref{eq:number-stripes}) allows for direct
quantitative comparison with data. The choice of
Eq.~(\ref{eq:freq_profile}) for the frequency profile comes from
phenomenological observations and it is not derived from the underlying
molecular interactions of the signalling gradients with PSM cells.
Nevertheless, our function for the frequency profile is well supported by
experimental data, see Fig.~\ref{fig:lewis.fit}.

It is important to note that a switch between modes can be achieved
while preserving the timing of somitogenesis by changing the shape
of the frequency profile: in Fig.~\ref{fig:phase.profile}f we show
results of simulations of a {\em mouse mode in zebrafish} with all
parameters given as in Table~I except for $\sigma$, which is 6 cell
diameters instead of 36 cell diameters, see also Supplementary
Movie~7. This implies that the number of stripes can change by
changing the shape of the frequency profile while leaving the collective
period and segment length unaffected.
%
%
Previous hypotheses for the different modes include changes in
period, loss of stripe specific cyclic gene enhancers, changes to
the stability of cyclic mRNA or different elongation velocities
\cite{gajewski06,holley07,elmasri04,gomez08}. The delayed coupling
theory indicates that changes to the frequency profile, potentially
through changes to FGF or Wnt signaling gradients in the PSM, and
different sizes of the PSM must be considered as well. This is
consistent with recent experiments reported in \cite{aulehla08},
where extra stripes of gene expression appear in a mutant with an
expanded PSM.


\subsection{Relation to regulatory network models}

Current regulatory network models for the genetic oscillations in
somitogenesis
\cite{jensen03,lewis03,monk03,cinquin07,rodriguez-gonzalez,goldbeter08},
undergo a Hopf bifurcation ---a generic mechanism by which
oscillations can appear in a dynamical system--- when varying some
parameters of the models, as for instance the transcriptional delays
\cite{bernard06,tiana07,feng07,verdugo08,momiji08}.
Although it is also valid in more general settings, our
Eq.~(\ref{eq:discrete}) can be obtained as the phase equation associated to the
normal form of a Hopf bifurcation when variations in the amplitude
of the oscillations can be neglected, and as such it can in principle be
derived from any of the dynamical systems associated with these
regulatory networks following standard procedures
\cite{hassard,kuramoto,nishii}. Hence our formulation represents a
simplification that captures general features and properties of more
detailed models.

The mechanism arresting the oscillations at the arrest front is a
different problem not addressed in our present work. While the
above mentioned models undergo a
Hopf bifurcation when varying one of their parameters, something
completely different (another kind of bifurcation triggered by the
variation of a different parameter of the models, for instance) may
be happening at the arrest front. The possibility that the
oscillations are coupled to a bistable switch related to the
signaling gradients in the PSM has been proposed
\cite{goldbeter07,santillan08}. In this scenario the arrest of the
oscillators would not be a result of the intrinsic mechanism of
the oscillations, but would result from an external signal.


\subsection{Implications of multistability}

Only stable solutions of our theory can be biologically relevant. In
addition, we hypothesize that unique solutions are required to
guarantee a robust behavior in the developing embryo. In the
presence of multiple stable solutions for the collective frequency,
fluctuations could drive the system to switch between these
different states, with dramatic consequences for healthy
development. For this reason, we conjecture that if several time
delays are consistent with a fit to experimental data, those
yielding a unique value of the collective frequency should be favored.
Biochemical evidence indicates that coupling time delays
should be relatively short compared to other signaling processes
in the vertebrate segmentation clock
\cite{heuss08}, thus likely precluding the observation of
multistability in such an embryonic system. Multistability has
been observed in other systems where coupling delays
can be large, with applications in biochemistry \cite{casagrande07},
chemistry \cite{kim01,manrubia04}, control theory \cite{beta04} or
laser physics \cite{wunsche05,franz08}, for example.
%

\subsection{Applications in somitogenesis and comparison to experiments}

Key quantitative experiments in vertebrate segmentation include
determination of segmentation rates \cite{schroeter07}, and the
analysis of expression patterns from {\em in situ} experiments
\cite{giudicelli07} and fluorescent reporter genes
\cite{masamizu06}. Our theoretical description allows for
quantitative analysis of these experiments.

The comparison to experimentally observed dynamic patterns of gene
expression permits the determination of the model parameters, which
are provided for wildtype zebrafish in Table~I. Future studies in
mutant embryos or embryos treated with different inhibitors will
reveal which parameters are affected. The parameters in our model
can be related to different cellular functions such as molecular
synthesis and trafficking of signals in the cell (coupling delay $\tau$); the strength
of intercellular signaling (coupling strength $\varepsilon$); the speed of a cell
autonomous oscillator (intrinsic frequency $\omega_L$); changes in the
signaling gradients responsible for the frequency profile
(decay length $\sigma$); and changes in the position of the arrest front
(reflected by the system length $L$). Thus, analysis of experimental
results using our theory can provide a deeper understanding of how
molecular changes lead to new phenotypes from the altered collective
dynamics of tissues.

Our framework can be extended to other developmental processes that
combine growth with a molecular clock. These are for instance
fore-limb autopod outgrowth and patterning \cite{pascoal07}, or
segmentation in short germ band insects, spiders, centipedes, and
other invertebrates that might form segments by a mechanism similar
to the one we described \cite{damens07,chipman08}.

\subsection{Summary}

The delayed coupling theory describes spatiotemporal patterns of
gene expression during morphogenesis in agreement with experimental
observations. Most importantly, our phenomenological theory provides
a unified quantitative framework relating the segmentation period
and cyclic patterns of gene expression to underlying properties, such
as the characteristics of intercellular communication, cell
autonomous oscillations, the spatial profile of the slowing of the
oscillators in the PSM, the rate of axial growth and the size of the
PSM. Our results indicate that
the specific spatial pattern of
cyclic gene expression in the PSM
does not affect the overall timing of
somitogenesis, but intercellular communication should be considered
as a fundamental mechanism in setting the collective frequency of the segmentation clock.\\

\section{Methods}

\subsection{Continuum limit}

Starting from Eq.~(\ref{eq:discrete-psm}) a continuum limit
describing the evolution of the phase can systematically be derived
for any value of the time delay. This continuum limit is valid when
the typical length scale of the modulations of the pattern is much
larger than the distance between oscillators, $a$. The
limit is obtained by letting the distance $a$
tend to zero, while the total number of oscillators $N$ tends to
infinite, in such a way that the length of the PSM, $L=Na$, remains
finite and constant. In the continum limit, we require a finite coupling strength
$\epsilon_c\equiv \lim_{a\to 0}\varepsilon/a^2$ to exist,
which implies that $\varepsilon$ scales as $a^2$.

The description based on discrete oscillators with phase
$\varphi_j(t)$ at a distance $a j$ from the arrest front (where $j$
is a discrete label) is substituted by a description defined in a
continuous field spanning from $x=0$ to $x=L$, where $x$ is a real
positive value giving the distance to the arrest front of a point of
the field with phase $\varphi(x,t)$. The resulting continuum
equation reads:
\begin{equation} \label{eq:continuum}
\dot{\varphi}(x,t) = \bar\omega(x) +v\nabla\varphi(x,t)+
\frac{\bar\epsilon_c}{2}\nabla^2\varphi(x,t),
\end{equation}
where $v$ is the velocity of the arrest front,
$\nabla$ denotes spatial derivatives
($\nabla=(\partial/\partial x)$ in one dimension,
$\nabla=(\partial/\partial x,\partial/\partial y)$
in two dimensions, and son on),
$\bar\omega(x)$ is a position dependent
effective frequency given by
\begin{equation} \label{eq:frequency.cont.approx.1}
\bar{\omega}(x) =\omega(x)\frac{1+2\pi m\epsilon_c/\omega_L}{1+\epsilon_c\tau},
\end{equation}
and $\bar\epsilon_c=\epsilon_c{(1+2\pi
m\epsilon_c/\omega_L)}/{(1+\epsilon_c\tau)}$ is the effective coupling
strength. The effect of the time delay appears through $\tau$ and
$m=[\tau\omega_L/2\pi]$, the nearest integer to $\tau\omega_L/2\pi$.
In analogy with $\omega_j$ in the discrete case, the intrinsic
frequency is defined as $\omega(x)=\omega_\infty(1-e^{-x/\sigma})$.
Note that for simplicity we have assumed that the intrinsic coupling
$\epsilon_c$ is constant throughout the PSM (as we did with
$\varepsilon$); it is straightforward to include a positional
dependence by substituting $\epsilon_c$ by $\epsilon_c(x)$ in all the
previous expressions.

We can simplify Eq.~(\ref{eq:continuum}) using the steady state
ansatz $\varphi(x,t)=\Omega t+\phi(x)$ as we did in the discrete
case:
\begin{equation} \label{eq:continuum.ansatz}
\Omega = \bar\omega(x) +v\nabla\phi(x)+
\frac{\bar\epsilon_c}{2}\nabla^2\phi(x).
\end{equation}
The boundary conditions for Eq.~(\ref{eq:continuum.ansatz}) are
$\nabla^2\phi(x)|_{x=0}=0$ and $\nabla\phi(x)|_{x=L}=0$. As in
the discrete case, we assume that the phase is defined and uniform
in the tailbud, $\phi(x>L)=\phi(L)$. This implies that at $x=L$ all
the derivatives in Eq.~(\ref{eq:continuum.ansatz}) vanish and
$\Omega=\bar\omega(L)$. In fact the right hand side of
Eq.~(\ref{eq:frequency.cont.approx.1}) coincides with the expression
for $\Omega$ obtained from solving  Eq.~(\ref{eq:Omega-epsilon-tau})
after linearization around values of the delay $\tau=2\pi
m/\omega_L$. In Fig~\ref{fig:omega.tau}a we show in red the
dependence of $\Omega=\bar\omega(L)$ with $\tau$ given by
Eq.~(\ref{eq:frequency.cont.approx.1}) for several values of $m$;
note that almost the whole range of stable solutions of
Eq.~(\ref{eq:Omega-epsilon-tau}) (solid blue) can be well
approximated by the continuum limit (red).

Eq.~(\ref{eq:continuum}) with $\bar\omega(x)$ given by
Eq.~(\ref{eq:frequency.cont.approx.1}) is valid when $\tau\approx
2\pi m/\omega_L$ for an integer $m$. A different equation can be
obtained in the cases where $\tau\approx 2\pi (m+1/2)/\omega_L$: it
corresponds to the continuum approximation of the unstable solutions
of Eq.~(\ref{eq:Omega-epsilon-tau}) shown by the broken blue lines
in Fig~\ref{fig:omega.tau}a.

Eq.~(\ref{eq:continuum.ansatz}) can be solved and the corresponding phase profile reads:
\begin{eqnarray}    \label{eq:solution.phase.nondimensional}
\phi(\xi) = {\nu} (1-\eta)^{-1} \, \big\{(1-\eta^2)  \\ \nonumber
-\mu \xi \left[e^{-\mu} -\eta e^{-\mu/\eta} \right] + \eta^2 e^{-\mu \xi / \eta} - e^{-\mu \xi} \big\}.
\end{eqnarray}
where we have defined the dimensionless coordinate $\xi= {x}/{L}$
and parameters $\mu={L}/{\sigma}$, $\nu ={\bar{\omega}_{\infty}
\sigma}/{v}$, and $\eta = {\bar{\epsilon_c}}/{2\sigma v}$; $\phi(0)$
has been set to $\phi(0)=0$ to fix an arbitrary constant.
Fig.~\ref{fig:phase.profile}a shows the shape of this phase profile.

The wavelength of the patterns of gene expression can be measured as
a function of the relative position $\xi$. In \cite{giudicelli07}
this is done experimentally, using a definition of the wavelength
$\lambda$ that in our notation can be expressed as the condition
$\phi(\xi+\lambda/2)-\phi(\xi-\lambda/2)=2\pi$. In the limit of
small coupling strength $\eta\ll 1$, we obtain a simple relation
between the local wavelength of the pattern $\lambda$ and the
position $\xi$ along the PSM given by Eq.~(\ref{eq:wavelength2})
of the main text.

\section*{Acknowledgments}
We thank Ingmar Riedel-Kruse, Lola Bajard, Ewa Paluch and the Oates
and J\"ulicher groups for enlightening discussions, and the MPI-CBG
fish facility for providing healthy fish. We also thank Julian Lewis
for sending us the raw experimental data from \cite{giudicelli07}
and for insightful comments on an earlier version of the manuscript.
LGM acknowledges support from CONICET. This work was supported by
the Max Planck Society.



\end{document}